\documentstyle[12pt,aaspp4,psfig,natbib]{article}

\newcommand{\vy}[2]{#1_{\scriptscriptstyle #2}}
\newcommand{\Ly}{Ly$\alpha$}

\def\gtorder{\mathrel{\raise.3ex\hbox{$>$}\mkern-14mu
             \lower0.6ex\hbox{$\sim$}}}
\def\ltorder{\mathrel{\raise.3ex\hbox{$<$}\mkern-14mu
             \lower0.6ex\hbox{$\sim$}}}
\def\proptwid{\mathrel{\raise.3ex\hbox{$\propto$}\mkern-14mu
             \lower0.6ex\hbox{$\sim$}}}
\def\0946{PG~0946+301}

\def\arcsec{\ifmmode '' \else $''$\fi}

\def\arcsecpoint{\ifmmode ''\!. \else $''\!.$\fi}

\def\kms{\ifmmode {\rm km\ s}^{-1} \else km s$^{-1}$\fi}
\def\Msun{\ifmmode {\rm M}_{\odot} \else M$_{\odot}$\fi}
\def\Lsun{\ifmmode {\rm L}_{\odot} \else L$_{\odot}$\fi}
\def\Zsun{\ifmmode {\rm Z}_{\odot} \else Z$_{\odot}$\fi}

\def\ergscm2{ergs\,s$^{-1}$\,cm$^{-2}$}
\def\icm3{{\rm cm}^{-3}}
\def\icm2{{\rm cm}^{-2}}
\def\qo{\ifmmode q_{\rm o} \else $q_{\rm o}$\fi}
\def\Ho{\ifmmode H_{\rm o} \else $H_{\rm o}$\fi}
\def\ho{\ifmmode h_{\rm o} \else $h_{\rm o}$\fi}

\def\vFWHM{\ifmmode v_{\mbox{\tiny FWHM}} \else
            $v_{\mbox{\tiny FWHM}}$\fi}
\def\CCF{\ifmmode F_{\it CCF} \else $F_{\it CCF}$\fi}
\def\ACF{\ifmmode F_{\it ACF} \else $F_{\it ACF}$\fi}
\def\Halpha{\ifmmode {\rm H}\alpha \else H$\alpha$\fi}
\def\Hbeta{\ifmmode {\rm H}\beta \else H$\beta$\fi}
\def\Hgamma{\ifmmode {\rm H}\gamma \else H$\gamma$\fi}
\def\Hdelta{\ifmmode {\rm H}\delta \else H$\delta$\fi}
\def\Lya{\ifmmode {\rm Ly}\alpha \else Ly$\alpha$\fi}
\def\Lyb{\ifmmode {\rm Ly}\beta \else Ly$\beta$\fi}
\def\Lyg{\ifmmode {\rm Ly}\beta \else Ly$\gamma$\fi}

\def\ciii{\ifmmode {\rm C}\,{\sc iii} \else C\,{\sc iii}\fi}
\def\civ{\ifmmode {\rm C}\,{\sc iv} \else C\,{\sc iv}\fi}

\def\nv{N\,{\sc v}}

\def\o5007{[O\,{\sc iii}]\,$\lambda5007$}

\def\ovi{O\,{\sc vi}}

\def\siiv{Si\,{\sc iv}}

\def\o{\o}
\begin{document}

\title{X-ray/UV CAMPAIGN ON THE MRK 279 AGN OUTFLOW: \\
       CONSTRAINING INHOMOGENEOUS ABSORBER MODELS 
}


\author{
 Nahum Arav\altaffilmark{1}, 
 Jelle  Kaastra\altaffilmark{2},
 Gerard~A. Kriss\altaffilmark{3}, 
 Kirk~T.~Korista\altaffilmark{4},
 Jack Gabel\altaffilmark{1},
 Daniel Proga\altaffilmark{5}
}

\altaffiltext{1}{CASA, University of Colorado, 389 UCB, Boulder, CO 80309-0389,
I:arav@colorado.edu}
 \altaffiltext{2}{SRON National Institute for Space Research
 Sorbonnelaan 2, 3584 CA Utrecht, The Nether\-lands}
 \altaffiltext{3}{STScI, Baltimore, MD 21218}
 \altaffiltext{4}{Western Michigan Univ., Dept.\ of Physics, 
 Kalamazoo, MI 49008-5252}
\altaffiltext{5}{JILA, University of Colorado, Boulder, CO 80309}


\begin{abstract}
We investigate the applicability of inhomogeneous absorber models in
the formation of AGN outflow absorption-troughs.  The models we
explore are limited to monotonic gradients of absorbing column
densities in front of a finite emission source.  Our main finding is
that simple power-law and gaussian distributions are hard pressed to
fit the Mrk~279 high-quality UV outflow data.  An acceptable fit for the
\ovi\ troughs can only be obtained by assuming unrealistic optical
depth values (upward of 100). The strongest constraints arise from the
attempt to fit the Lyman series troughs.  In this case it is evident
that even allowing for complete freedom of both the power-law exponent
and the optical depth as a function of velocity cannot yield an
acceptable fit.  In contrast, partial covering models do yield good
fits for the Lyman series troughs.  We conclude that monotonic
inhomogeneous absorber models that do not include a sharp edge in the
optical depth distribution across the source are not an adequate
physical model to explain the trough formation mechanism for the
outflow observed in Mrk~279.

\end{abstract}

\section{INTRODUCTION}

AGN outflows are evident by resonance line absorption troughs, which
are blueshifted with respect to the systemic redshift of their
emission counterparts. In Seyfert galaxies velocities of several
hundred \kms\ (Crenshaw et~al.\ 1999; Kriss et~al.\ 2000) are
typically observed in both UV resonance lines (e.g.,
\civ~$\lambda\lambda$1548.20,1550.77,
\nv~$\lambda\lambda$1238.82,1242.80,
\ovi~$\lambda\lambda$1031.93,1037.62 and \Ly), as well as in X-ray
resonance lines (Kaastra et~al.\ 2000, 2002; Kaspi et~al.\ 2000, 2002). Similar
outflows (often with significantly higher velocities) are seen in
quasars which are the luminous relatives of Seyfert galaxies (Weymann
et~al.\ 1991; Korista, Voit, Morris, \& Weymann 1993; Arav et~al.\
2001a). Reliable measurement of the absorption column densities in the
troughs is crucial for determining the ionization equilibrium and
abundances of the outflows, and the relationship between the UV and
the ionized X-ray absorbers.

In the last few years our group (Arav 1997; Arav et~al.\ 1999a; Arav
et~al.\ 1999b; de~Kool et~al.\ 2001; Arav et~al.\ 2002, 2003) and
others (Barlow 1997, Telfer et~al.\ 1998, Churchill et~al.\ 1999,
Ganguly et~al.\ 1999) have shown that in quasar outflows most lines
are saturated even when not black.  As a consequence the apparent
optical depth method, which stipulates that the optical depth
$\tau_{ap}\equiv-\ln(I)$, where $I$ is the residual intensity in the
trough, is not a good approximation for outflow troughs. In addition
using the doublet method (Barlow 1997, Hamann et al 1997) we have also
shown that in many cases the shapes of the troughs are almost entirely
due to changes in the line of sight covering as a function of
velocity, rather than to differences in optical depth (Arav et~al.\
1999b; de~Kool et~al.\ 2001; Arav et~al.\ 2001a). Gabel et al.\ (2003)
show the same effect in the outflow troughs of NGC~3783, as does Scott
et al.\ (2004) for Mrk~279.  As a consequence, the column densities
inferred from the depths of the troughs are only lower limits.

Most of the analysis cited above relies on comparing two troughs
arising from the doublet lines of a given ion (e.g., the \civ, \nv\
and \ovi, doublets mentioned above).  Based on the oscillator strength
of the doublet components, the blue transition must have twice the
optical depth as the red transition.  When the apparent optical depth
 method is used on fully resolved, high S/N, non-blended
doublet troughs, the ratio of their $\tau_{ap}$ is rarely 1:2.  In
most cases, the trough from the red doublet has more than half of the
blue $\tau_{ap}$, up to equality in many cases.  In order to determine
the actual optical depth (and hence column density), we use a pure
partial covering model.  We assume that only a fraction $C$ of the
emission source is covered by the absorber and then solve for a
combination of $C$ and $\tau$ that will fit the data of both doublet
troughs while maintaining the intrinsic 1:2 optical depth ratio
(see \S~3 for the full formalism, including velocity dependence).
Implicit in this model are the assumptions that the absorber covering
the $C$ fraction of the source has the same optical depth across this
area, and that the rest of the source is covered by material with
$\tau=0$ in that transition.

Although widely used, the validity of the pure partial covering model
should be questioned on two fronts.  First, mathematically we solve
for two unknowns ($C$ and $\tau$) given two residual intensity
equations.  As long as the ratio of the residual intensities is in the
permitted physical range (see \S~3), such a procedure will always
yield a solution.  A principal issue is, how valid is this
solution physically?  To test the partial covering model we need more
lines from the same ions and efforts in that direction will be
described in forthcoming papers (Gabel et al.\ 2004, Arav et al.\
2004).  The second question has to do with the rigid distribution
(essentially a step-function) of absorbing material assumed by the
partial covering model.  Here one may ask if an inhomogeneous
distribution of absorbing material across the emission source can
yield a good alternative to the partial covering model.

The first attempt to study generic inhomogeneous distributions of
absorbing material in AGN outflows was done by de Kool, Korista \&
Arav (2002, hereafter dKKA).  A formalism was developed to simulate
the effects of an inhomogeneous absorber on the emerging spectrum and
several examples of fitting existing data with this model were
presented.  In this paper we begin by giving a simplified version of
the dKKA formalism, and explain the spectrum arising from simple
monotonic material distributions (\S~2).  Since little constraints can
be put on the formation of a single trough, in \S~3 we simulate and
explain the formation of doublet troughs.  In \S~4 we use these
simulations in an attempt to fit the \ovi, \nv\ and Lyman series
troughs seen in high-quality observations of Mrk~279.  In light of
these results, we assess the physical viability of inhomogeneous
absorber models in \S~5


\section{MONOTONIC OPTICAL DEPTH DISTRIBUTIONS}

In this section we introduce the formalism needed to calculate 
the attenuation caused by an inhomogeneous absorbing medium.
Our starting point is equation (6) from dKKA:

\begin{equation} 
F(\lambda)=\int\int S(x,y,\lambda)e^{-\tau(x,y,\lambda)}dxdy,
\label{eq:general}
\end{equation}
where $S(x,y,\lambda)$ is the surface brightness distribution of the
background source and $\tau(x,y,\lambda)$ is the line of sight optical
depth at wavelength $\lambda$ in front of a specific $(x,y)$ location,
as defined by dKKA equations (3-5).  We find that the
formalism that depends on real spatial variables $(x,y)$ is simpler and
more intuitive than working in $\tau$ space as was originally done in
dKKA (their equations 7-10, 13 and 15). 

In its most general form, Equation (\ref{eq:general}) is not practical
for modeling spectra.  Both $S(x,y,\lambda)$ and $\tau(x,y,\lambda)$
each have two functional degrees of freedom for a given $\lambda$.  We
need a  simpler and testable model.  To this end, we introduce the
following simplifying assumptions and comment about their physical
validity and plausibility.

\begin{enumerate}
\item $S(x,y,\lambda)=1$

This is probably the weakest of our simplifications.  For an accretion
disk, it is probable that $S(x,y)$ is a strong function of distance
from the central source.  Moreover, there are at least two main
emission sources at the wavelength range of the absorption troughs:
Continuum emission and broad emission line flux.  These differ from
each other greatly by their physical size and the emissivity as a
function of wavelength (see Gabel et~al.\ 2004). Nonetheless, it is
essential to at least start with $S(x,y,\lambda)=1$ models in order to
gain insight about absorption-trough formation in inhomogeneous media.

\item $\tau(x,y,\lambda)=\tau(x,\lambda)$

This simplifying assumption entails  little loss of generality.
We can think of $e^{-\tau(x,\lambda)}$ as the integrated attenuated
flux along a $y$ strip at a specific $(x,\lambda)$.

\item $\tau(x,\lambda)=\tau_{\rm{max}}(\lambda)x^a$\ \ \ \ \ \ or \ \ \ \ \ \  
$\tau(x,\lambda)=\tau_{\rm{max}}(\lambda)e^{-(x/b)^2}$

Since there are infinite functional possibilities for the $\tau(x)$
distribution, we limit our discussion to two simple and well-behaved
functional forms.  Each of these functions depends on
two free parameters.  We also discuss a two parameter step function,
which is the mathematical representation of the pure partial covering
model.

\end{enumerate}

\subsection{Power-Law, Gaussian and Step-function Distributions}

Using the above simplifications, we begin by examining simple
monotonic distributions of optical depth in front of a one dimensional
emission source.  By construction, the emission source spans the
interval $0<x<1$ and we assume that the surface brightness distribution
is constant over this interval and given by $S(x,y,\lambda)=1$ (see
discussion above). Finally, for simplicity we ignore the $\lambda$ (or
velocity) dependence of both $S$ and $\tau$. We will examine the role
of velocity dependence in \S~3.  To gain better understanding of the
absorption behavior of such a medium we define the following
parameters:

\begin{enumerate}
\item $\overline{\tau}\equiv\int_0^1\tau(x)dx$,
 the average  optical depth in front
of the emission source.  This is the relevant quantity for computing
and comparing column densities ($\rm{N}_{\rm{ion}}$).

\item $I\equiv\int_0^1e^{-\tau(x)}dx$, the residual intensity of the
transmitted flux (normalized to 1 in the absence of absorption) that
an observer far from the absorber will measure.

\item $\tau_{ap}\equiv -\ln(I)$,  apparent optical depth, which is the
optical depth inferred from $I$ assuming a uniform absorbing slab
that fully covers the emission source.  This is the general case for
ISM and IGM absorbers.

\item $T$, transmission parameter, measures the fraction of the
emission source that is covered by $\tau<0.5$.  As we show below, for
steep $\tau(x)$ distributions $I\simeq T$. Even more accurately,
$\Delta I\simeq\Delta T$ for two similar distributions with different
$\overline{\tau}$.  


\end{enumerate}

\begin{figure}[ht]
\centerline{\psfig{file=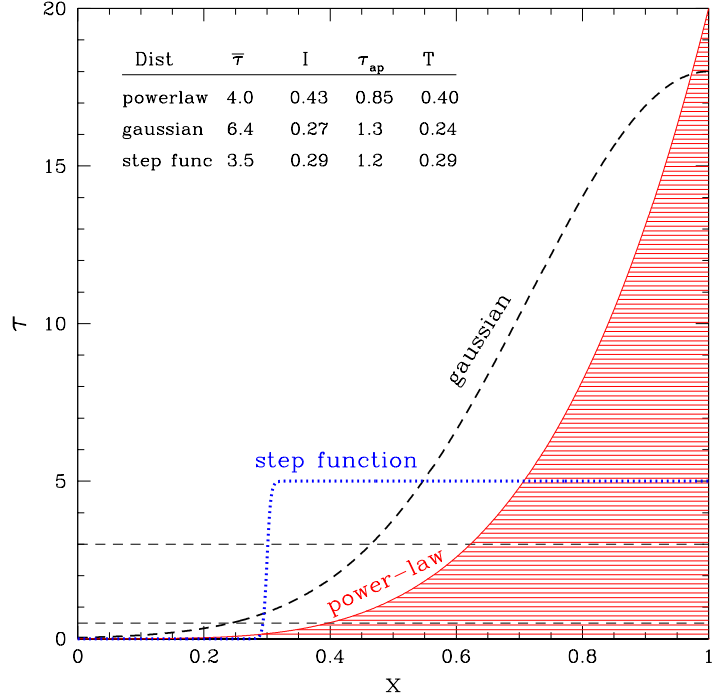,angle=0,height=10.0cm,width=10.0cm}}
\caption{Several monotonic distributions of absorbing material in front
of an emission source.  The constant emission source coincides with
the $x$ axis, and the observer is far above the plot.  For the
power-law case we shaded the region of absorbing material, and for all
distributions the absorbing material is below  the curve. For each
distribution the optical depth in front of an $x$ location is given
by: \newline Power-law: $\tau(x)=\tau_{\rm{max}}x^a$, \ \
$\tau_{\rm{max}}=20, \ \ a=4$ 
\newline Gaussian:
$\tau(x)=\tau_{\rm{max}}e^{-[(x-1)/b]^2}$, \ \ $\tau_{\rm{max}}=18, \
\ b=0.4$
\newline
 Step function:
$0<x<\vy{x}{0}\ \ \tau(x)=0,\ \ \ \vy{x}{0}<x<1 \ \ \tau(x)=\tau_{\rm{max}}$; \ \ $\tau_{\rm{max}}=5, \ \ \vy{x}{0}=0.3$
\newline
Two dashed lines bracket the interval $0.5<\tau<3$.  As can be judged
from the inset table, the residual intensity ($I$) is mainly decided
by the extent of the $\tau<0.5$ absorber ($T$). For any distribution,
material with $\tau>3$ has negligible effect on the trough since less
than 5\% $(e^{-3})$ of the flux at any given point is being
transmitted.}
\label{fig:tau_dist}
\end{figure}

In figure \ref{fig:tau_dist} we illustrate several monotonic absorbing-material
 distributions.  
The inset table displays some of the parameters
defined above that are associated with these distributions.
For calculating the transmitted residual flux $I$,
we divided the emission zone into 1000 elements, calculated $I$ at
each one according to the value of $\tau$ in front of it, summed all
the contributions and normalized by the number of elements.   To aid the
visualization of such distributions we make the assumption that the
number density of the absorbing ion $n_{\rm{ion}}$ is constant.  This
allows us to create simple geometrical illustrations of the
absorbing-matter distribution, where the optical depth axis is
proportional to the spatial depth of the absorber.  We note that this
assumption does not change the calculation of emergent flux, but in
general will affect the conversion of optical depth to total column
density.

Following dKKA, for the power-law case we assume that
$\tau_{\rm{min}}=0$, which reduces the number of free parameters to
two: $\tau_{\rm{max}}$, the maximum optical depth in the distribution,
and $a$, the power-law exponent.  Although for power-law distributions
dKKA give closed form solutions for equation (\ref{eq:general}), these
are expressed as the product of gamma functions and incomplete gamma
functions (see their eq. 13), which are not particularly illuminating.
We therefore rely solely on numerical results and attempt to clarify
their origin and significance.  To ease comparison with the power law
case, the actual gaussian distribution plotted is
$\tau(x)=\tau_{\rm{max}}e^{-[(x-1)/b]^2}$, with no loss of generality.

The amount of material in each distribution is given by
$\overline{\tau}$, and the ratio $\overline{\tau}/\tau_{ap}$ can be
interpreted as the amount of material ``hidden'' in the inhomogeneous
absorber compared to what we naively deduce by using $\tau_{ap}$.  As
we show with real data (see \S~4), this ratio can become very large
when attempting to fit doublet troughs with power-laws and gaussian
distributions.  For all plotted distributions we note the similarity
in value between $I$ and $T$.
The  plotted  step function shows the 
inhomogeneous absorber representation of a pure partial covering. 

\clearpage

\clearpage

\section{FORMATION OF DOUBLET TROUGHS}

Depending on which model we use for the absorber, the residual
intensity in a single absorption trough can yield vastly different
estimates for the optical depth that causes it.  We know that the
apparent optical depth method, which works so well for the ISM and IGM
absorbers, is a very poor model for many AGN outflows troughs, where
$\tau_{ap}\equiv -\log(I)$ gives only a lower limit for the real optical
depth.  This is evident when more troughs from the same ion are added
as constraints on our absorbing model.  Several doublet lines commonly
produce absorption troughs in the spectrum of AGN outflows, the most
notable are \civ~$\lambda\lambda$1548.20,1550.77,
\nv~$\lambda\lambda$1238.82,1242.80 and
\ovi~$\lambda\lambda$1031.93,1037.62.  These lines provide sufficient
constraints to test the apparent optical depth method.  For all the
above doublets, the oscillator strength of the blue (the shorter
wavelength) transition is exactly twice that of the red transition.
Therefore, the expected optical depth ratio is almost exactly 2:1 in
favor of the blue transition. (It is not exactly 2:1 since the
wavelengths differ by a small amount.)  For the apparent optical
depth method to hold, we must have $\tau_{ap}$(blue)=$2\tau_{ap}$(red)
which requires: $I_B=I_R^2$, where $I_B$ and $I_R$ are the residual
intensities of the blue and red absorption troughs, respectively.  For
ISM and IGM absorber this relationship holds to an excellent degree
(Arav et~al.~2001a). In contrast, in most AGN outflow troughs we find
$I_B\neq I_R^2$, instead we detect the whole range $I_R\geq I_B\geq
I_R^2$.  (Values outside this range are unphysical and usually can be
attributed to the errors associated with the data [Hamann et~al.\ 1997].)

In order to explain the observed $I_R\geq I_B\geq I_R^2$, the standard
method is to invoke partial covering of the emission source.  In its
simplified form, we assume that the absorber covers only a portion $C$
(which generaly depends on the velocity) of a constant emission source
(at a given $\lambda$).  We further assume that the $\tau$
distribution in front of the covered part of the source is constant.
With these assumptions, the absorption equations are:
\begin{equation}
I_R(v)-(1-C(v))  = C(v)e^{-\tau(v)} 
\label{eq:covering1}
\end{equation}
\begin{equation}
 I_B(v)-(1-C(v))  = C(v)e^{-2\tau(v)},
\label{eq:covering2}
\end{equation}
where $v$ is the velocity of the outflow, $C(v)$ is the effective
covering fraction (see Arav et al. 1999b), and $\tau(v)$ is the
optical depth of the red doublet component.  From these two equations
we obtain:
\begin{equation}
C(v) = {{I_R(v)^2 - 2I_R(v) + 1}\over{I_B(v) - 2I_R(v) +1}} , \ \ \ \ \ \
\tau(v)=-\ln\left({I_R(v)-I_B(v)}\over{1-I_R(v)}\right),
\label{eq:doublet}
\end{equation}
By construction, the partial covering model will always give a  
$C(v)$ and $\tau(v)$ solution to equations 
(\ref{eq:covering1}) and  (\ref{eq:covering2}), provided that
$I_R\geq I_B\geq I_R^2$. In order to test the validity of the model
we must work with more than two lines from the same ion (see Gabel et~al.\
2004).  

For the inhomogeneous absorber we need to test if, and under what
conditions, the absorber satisfies the constraints imposed by
doublet lines.  In doing so we make use of and expand the simulation
results derived by dKKA, followed in the next section by applying
these results to data of Mrk 279.  To simulate doublet
data, we start by computing the residual intensity for a given
power-law distribution $\tau(x)=\tau_{\rm{max}}x^a$, thus obtaining
$I(\tau_{\rm{max}},a)$.  We define the red doublet residual intensity
to be $I_R\equiv I(\tau_{\rm{max}},a)$ and obtain the blue doublet
residual intensity, by using twice the $\tau_{\rm{max}}$ value for the
same distribution, $I_B=I(2\tau_{\rm{max}},a)$.  To obtain a meaningful
comparison between different power-law ($a$) distributions we convert
the resulting $I(\tau_{\rm{max}},a)$ to $I(\overline{\tau},a)$.
Apart from the last step, this is the same procedure dKKA used for producing
their figure (2).  However, unlike dKKA, we begin with  looking at a single
velocity position.  This is done 
in order to gain better understanding 
for the formation of doublet troughs from inhomogeneous absorbers.
We discuss the velocity dependency in the next section.
Following the power-law simulation we repeated the same procedure to produce 
doublet simulations from gaussian distributions.

\begin{figure}[ht]
\centerline{\psfig{file=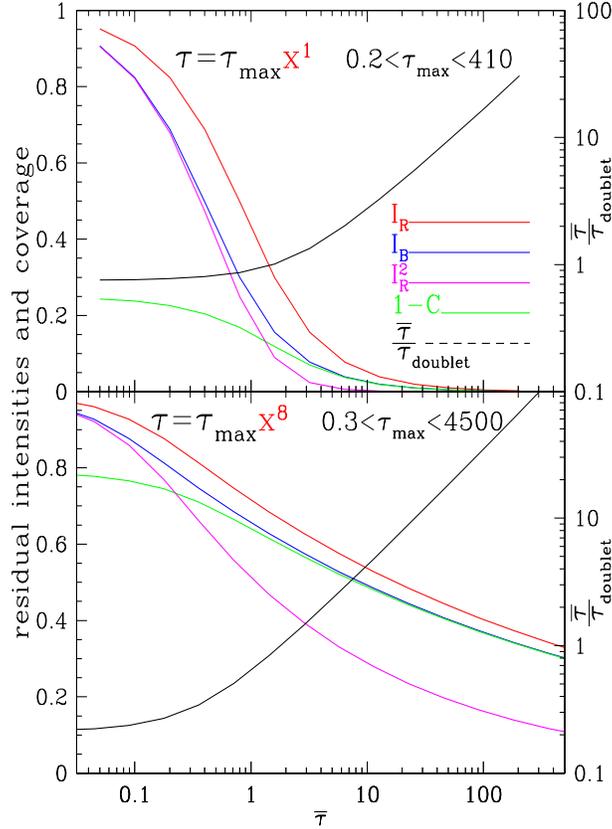,angle=0,height=12.0cm,width=8.0cm}}
\caption{Formation of doublet troughs from power-law distributions of
absorbing material (given at the top of each panel).  For both
distributions we display the
following results: red, residual intensity of the red doublet component
($I_R$); blue, residual intensity of the blue doublet component
($I_B$), which is created by doubling the optical depth used to
simulate $I_R$; magenta, $I^2_R$, what $I_B$ would have been for complete 
coverage homogeneous absorber; 
green, $1-C$ where $C$ is the covering factor one
would derive using the pure partial covering equations
(\ref{eq:doublet}) for the simulated $I_R$ and $I_B$; black
$\overline{\tau}/\tau_{\rm{doublet}}$ (read from the right axis),
where $\tau_{\rm{doublet}}$ is the optical depth derived by using the
pure partial covering equations (\ref{eq:doublet}).}
\label{fig:tau_doublet_tau_av}
\end{figure}

Figure \ref{fig:tau_doublet_tau_av} shows the simulated results for two
power law distributions, $a=1$ and $a=8$.  The simulated $I_R$ and
$I_B$ are displayed as well as $I^2_R$.  As discussed above, for a
given $I_R$ the allowed physical range of  $I_B$ is always
between  $I_R$ and $I^2_R$, and  it is important to keep in mind the
narrowness of the physically allowed range when comparing power law
fits to actual doublet data.  Also shown is $1-C$ where $C$ is the
covering factor one would derive using the pure partial covering
equations (\ref{eq:doublet}) for the simulated $I_R$ and $I_B$.  A
common result from analyzing real data is that the $1-C$ curve is
almost identical with $I_B$.  In the simulations, we observe that for
large enough $\overline{\tau}$ (roughly above 3 in both cases), $1-C$
is indeed almost identical with $I_B$.  As was found by dKKA, we
confirm that for a shallow power-law ($a=1$), $1-C$ is consistent with
$I_B$ only for $0<I_B< 0.1$, and for a steep power-law ($a=8$) $1-C$ is
consistent with $I_B$ for $0<I_B<0.6$.  However, in the steep power-law
case, it is difficult to obtain low values of $I_B$. In the example
shown, $I_B=0.35$ only at $\overline{\tau}=300$.  The dashed curves
show the ratio of $\overline{\tau}/\tau_{\rm{doublet}}$ (read from the
right axis), where $\tau_{\rm{doublet}}$ is the optical depth derived
by using the pure partial covering equations (\ref{eq:doublet}).  This
ratio is a good indicator for the amount of ``hidden'' column density
in the inhomogeneous absorber compared to what we deduce from the pure
partial covering model. We note that $\tau_{\rm{doublet}}\geq \tau_{ap}$, 
therefore $\overline{\tau}/ \tau_{ap}$ will be  correspondingly larger. 

Let us look at the mechanism which creates the simulated $I_R$ and
$I_B$ for the steep power-law ($a=8$).  This case is interesting
since for a given $\tau$, $I_R$ and $I_B$ are quite close in value and
$I_R-I_B$ is considerably smaller than $I_R-I_R^2$ or $I_B-I_R^2$.
Such a situation is often seen in real outflow data.  The driver
behind the finite yet small $I_R-I_B$ is closely related to the area
of the absorber that is covered by small optical depth.  For this
purpose we defined the transmission parameter $T$, which measures the
fraction of the emission source that is covered by $\tau<0.5$.  We
find that for steep power-laws, $T$ is almost identical with the
residual intensity.  For example, for $\overline{\tau}=50, I=0.41,
T=0.40$ and for $\overline{\tau}=100, I=0.37, T=0.36$.  Since the
ratio of $\overline{\tau}$ in these two examples is 2, the residual
intensities can be taken as $I_B(\overline{\tau}=100)$ and
$I_R(\overline{\tau}=50)$ (which is how figure
\ref{fig:tau_doublet_tau_av} was made).  $I_R-I_B$ is almost fully
attributed to $\Delta T$, which suggests a simple geometrical
interpretation: The residual intensities are mainly determined by the
optically thin ($0.5>\tau$) extent of the absorber over the emission
source ($T$), and the difference in doublet intensities, $I_R-I_B$, is
almost entirely explained by the $\Delta T$ of the two distributions.

\begin{figure}[ht]
\centerline{\psfig{file=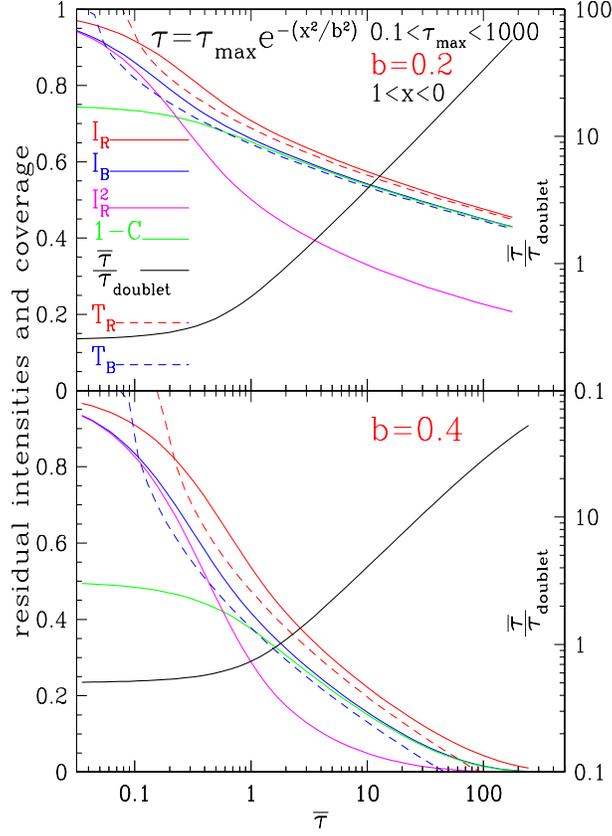,angle=0,height=12.0cm,width=8.0cm}}
\caption{Formation of doublet troughs from gaussian distributions of
absorbing material (see top panel).  The presentation is similar to
that of Fig.~\ref{fig:tau_doublet_tau_av} with two additional dashed
curves, which show the transmission parameter for the gaussian
distributions: T$_R$  and T$_B$ are for the distribution that
produces I$_R$ and I$_B$, respectively. For the narrow $b=0.2$
gaussian it is evident that I$_R\simeq$T$_R$ and
I$_B\simeq$T$_B$ (as is the case for steep power-law
distributions).  Even for broader gaussians ($b=0.4$) this
approximation is good to within 10\% -- 20\% across most of the
$\overline{\tau}$ range.}
\label{fig:tau_doublet_tau_av_gaussian}
\end{figure}

Figure \ref{fig:tau_doublet_tau_av_gaussian} shows the simulated
results for two gaussian distributions, $b=0.4$ and $b=0.2$.  The
additional dashed curves give $T_R$ and $T_B$ the transmission
parameter for the red and blue troughs respectively.  It is evident
that I$_R\simeq$T$_R$ and I$_B\simeq$T$_B$ (as is the case for steep
power-law distributions).  Even for broader gaussians ($b=0.4$) this
approximation is good to within 10\% -- 20\% across most of the
$\overline{\tau}$ range. Comparing figures
\ref{fig:tau_doublet_tau_av} and \ref{fig:tau_doublet_tau_av_gaussian}
we see that there is little qualitative difference between the
troughs  produced by power-laws and gaussians.  This is expected since 
the behavior of both functions in the $\tau<3$ region is quite similar 
for steep and high $\tau_{\rm{max}}$ cases (see Fig.~\ref{fig:tau_dist}).
Since the results of power-law and gaussian models are very similar,
we will concentrate on power-laws in the rest of this paper.

Having $I_R-I_B$ small but finite at high $\overline{\tau}$ is both a
strength and a weakness for the inhomogeneous absorber model.  In real
high-quality data, we often see occurrences that look like the lower
panel of figure \ref{fig:tau_doublet_tau_av}.  That is, $I_B$ is close
to, but distinctively smaller than $I_R$ while $I_B$ is considerably
larger than $I_R^2$ (the homogeneous full coverage absorber
model). These occurrences are statistically difficult to explain using
the pure partial coverage model (eqs.  \ref{eq:covering1} and
\ref{eq:covering2}), where for $\tau_{red}>3$, $I_R-I_B\simeq0$.  We are
therefore left with the question of why in so many outflow absorbers,
mother nature chose optical depth $0.5<\tau_{red}<3$ (Below
$\tau_{red}=0.5$ it is difficult to distinguish between the homogeneous full
coverage absorber and the pure partial coverage models.).  As shown in
figure \ref{fig:tau_doublet_tau_av}, inhomogeneous absorber models can
produce a small but finite $I_R-I_B$  at much higher $\overline{\tau}$
values.  The weakness is related to the observed examples of
$I_R-I_B\simeq0$, which of course reverses the
argument. $I_R-I_B\simeq0$ is naturally explained as $\tau_{red}>3$
occurrences for the pure partial coverage model.  However, the
inhomogeneous power-law absorber models maintain a finite $I_R-I_B$ even
at absurdly high values of $\overline{\tau}$.  This point is
strengthened by the fitting of \ovi\ doublet troughs in the Mrk 279
data (\S~5.2).  A related weakness of inhomogeneous power-law absorber models
is their inability to create small values of $I_R, I_B$ for steep 
distributions (e.g., the $a=8$ example in fig. \ref{fig:tau_doublet_tau_av})
and reasonable values of  $\overline{\tau}$.  Shallow power-law distributions
can do so but at the cost of having unrealistically large $I_R-I_B$ values.
In the following section we will test these conjectures on real outflow
data.

\clearpage

\clearpage

\section{FITTING REAL DATA }

On May 2003 we obtained simultaneous X-ray and UV observations
of the Mrk~279 AGN outflow. (Description of the UV observations is found in 
Gabel et~al.\ 2004, and the X-ray observations in Costantini et~al.\ 2004.)
 The 92 ksec FUSE data yielded
the highest quality \ovi\ trough-spectrum of any AGN outflow to date,
and the 16 orbits HST/STIS/E140M observation yielded high 
quality \nv\ troughs. The combined HST/FUSE spectrum gives 
high signal-to-noise \Lya, \Lyb\ and \Lyg\ troughs.
In this section we use the inhomogeneous absorber formalism
developed in the previous section to fit these data and check whether 
these simple distributions are an adequate model for real outflow troughs.

To fit the \ovi\ and \nv\ troughs with inhomogeneous absorber models we
start by normalizing the data, which is done by dividing the observed
flux by an unabsorbed emission model.  In doing so we implicitly
assume that the inhomogeneous absorber has the same characteristics in
front of both the continuum and BEL sources.  We then fit the red
doublet data at each velocity bin as follows: we choose a power law
value and using the formalism described in \S~2 iterate until a
specific $\tau_{\rm{max}}$ gives us the observed value of $I_R$.  We
use the $\overline{\tau}$ of this distribution when comparing with the
optical depth $\tau_{\rm{doublet}}$ derived from the doublet equations
(\ref{eq:doublet}).  Following the procedure described in \S~3, we
double the fitted $\tau_{\rm{max}}$ and take the residual intensity
from the same distribution with $2\tau_{\rm{max}}$ as $I_B$.  This
process is repeated for all the velocity bins in the trough.  If the
simulated $I_B$ curve for any given power-law index gives a good fit
for the entire observed $I_B$ trough, we will have a strong indication
that the material is indeed inhomogeneously distributed, with a
specific power law.  A similar fitting procedure is used for the Lyman
series troughs.

\subsection{\ovi\ Troughs}

\begin{figure}
\centerline{\psfig{file=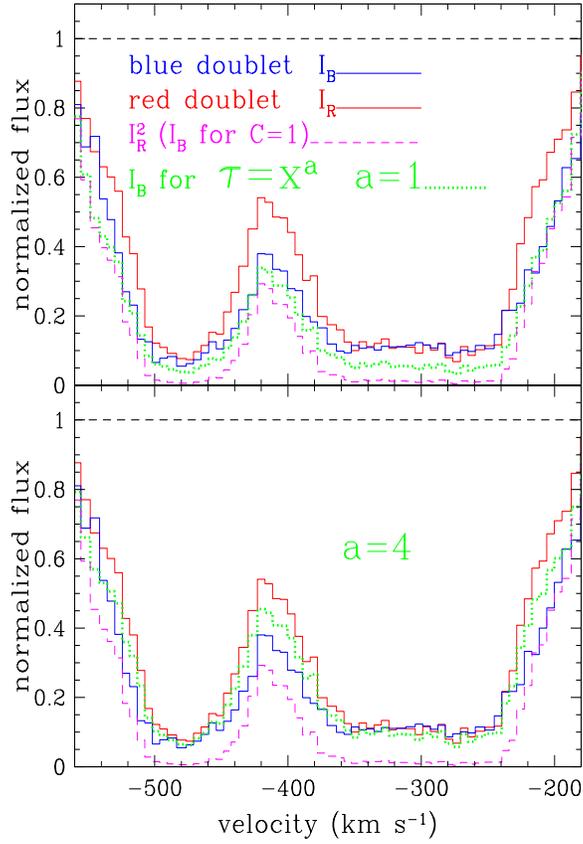,angle=0,height=12.0cm,width=8.0cm}}
\caption{Power-law fits to the \ovi\ outflow troughs seen in Mrk~279
(2003, FUSE spectrum).  The normalized data for the red and blue
doublet troughs are plotted, as well as $I^2_R$, which is the expected
value of $I_B$ for complete coverage homogeneous absorber.  The
physically allowed range of $I_B$ for any material distribution is
between the $I_R$ and $I^2_R$ curves. For the chosen power-law model
we show the predicted $I_B$ from the same distribution that gave an
exact match to the observed $I_R$ .}
\label{fig_o6_a1_a4}
\end{figure}

\begin{figure}
\centerline{\psfig{file=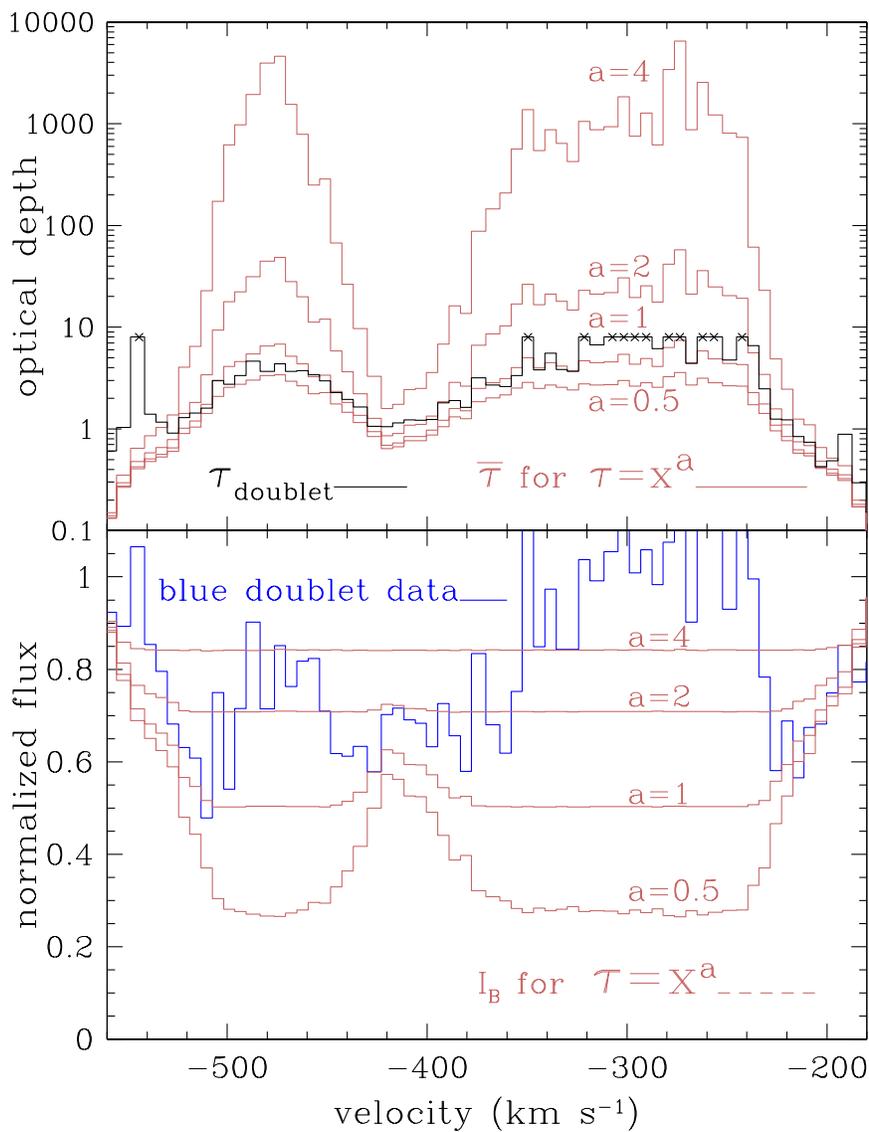,angle=0,height=16.0cm,width=12.0cm}}
\caption{Four power-law fits for the Mrk~279 \ovi\ outflow troughs.
The data shown in Fig.~\ref{fig_o6_a1_a4} were renormalized by dividing
all the curves by $I_R$.  This allows us to compare several solutions and
examine the behavior of $I_B/I_R$.  In this presentation, it is
clear that none of the $I_B$ solutions fits the blue doublet data
well. A combination of power-laws with $1<a<10$ is needed to fit the
full blue trough.}
\label{fig_o6_renorm}
\end{figure}

In figure \ref{fig_o6_a1_a4} we show the \ovi\ fitting results for two
power-law models.  It is clear that the $a=1$ case gives a poor fit to
the blue doublet-component data for most of the velocity range $-500<v<-230$ \kms. This
is mainly due to the inability of a shallow power law to produce
$I_B\sim I_R$ (see Fig.~\ref{fig:tau_doublet_tau_av}). A steeper
power-law ($a=4$) does a much better job in fitting the $I_B\sim I_R$
parts of the trough, but cannot reproduce the regions where $I_B$ is
roughly half way between $I_R$ and $I_R^2$.  The general behavior is
shown in the lower panel of figure \ref{fig_o6_renorm}, where we
re-normalize the \ovi\ data by dividing all the curves by $I_R$.  In
this presentation $I_R=1$ and the behavior of $I_B$ with respect to
$I_R$ is made clearer.  Moreover, we can now examine four power-law
models and better quantify their relationship to $I_B$ and $I_R$. We
observe that the ratio $I_B/I_R$ is not constant and shows
considerable structure. In contrast, for steep power-laws that give
better overall fits to $I_B$, the ratio of simulated $I_B/I_R$ is
constant.  These contrasting behaviors explain why steep power-law (or
narrow gaussian) models do not provide a good fit for the entire $I_B$
trough.

However, the main problem with steep power-law (or narrow gaussian)
models is the unrealistic high value of $\overline{\tau}$ necessary to
obtain the observed $I_R$.  This issue was already discussed in the
simulation of doublet data (\S~3).  In the upper panel of figure
\ref{fig_o6_renorm} we show the $\overline{\tau}$ curves for the models
shown in figure \ref{fig_o6_a1_a4} and the lower panel of
\ref{fig_o6_renorm}.  The steep power-law models ($a=2,4$), which
provide a better fit for most of the observed trough, necessitate high
$\overline{\tau}$.  This behavior is  an artifact of the need to
have a steep power-law model with a small $T$ parameter.  
We conclude
that simple power-law and gaussian distributions are hard pressed to
fit the Mrk~279 high-quality \ovi\ trough data.

\clearpage

\clearpage

\subsection{\nv\ Troughs}

\begin{figure}
\centerline{\psfig{file=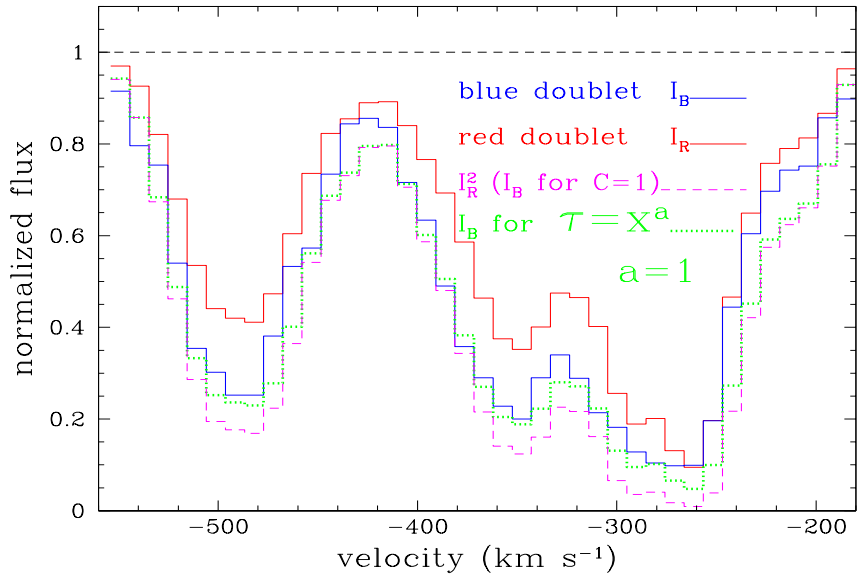,angle=-90,height=8.0cm,width=12.0cm}}
\caption{Power-law fit to the \nv\ outflow troughs seen in Mrk~279
(2003, HST/STIS spectrum), same presentation as Fig.~\ref{fig_o6_a1_a4}.}
\label{fig_n5_a1}
\end{figure}

\begin{figure}
\centerline{\psfig{file=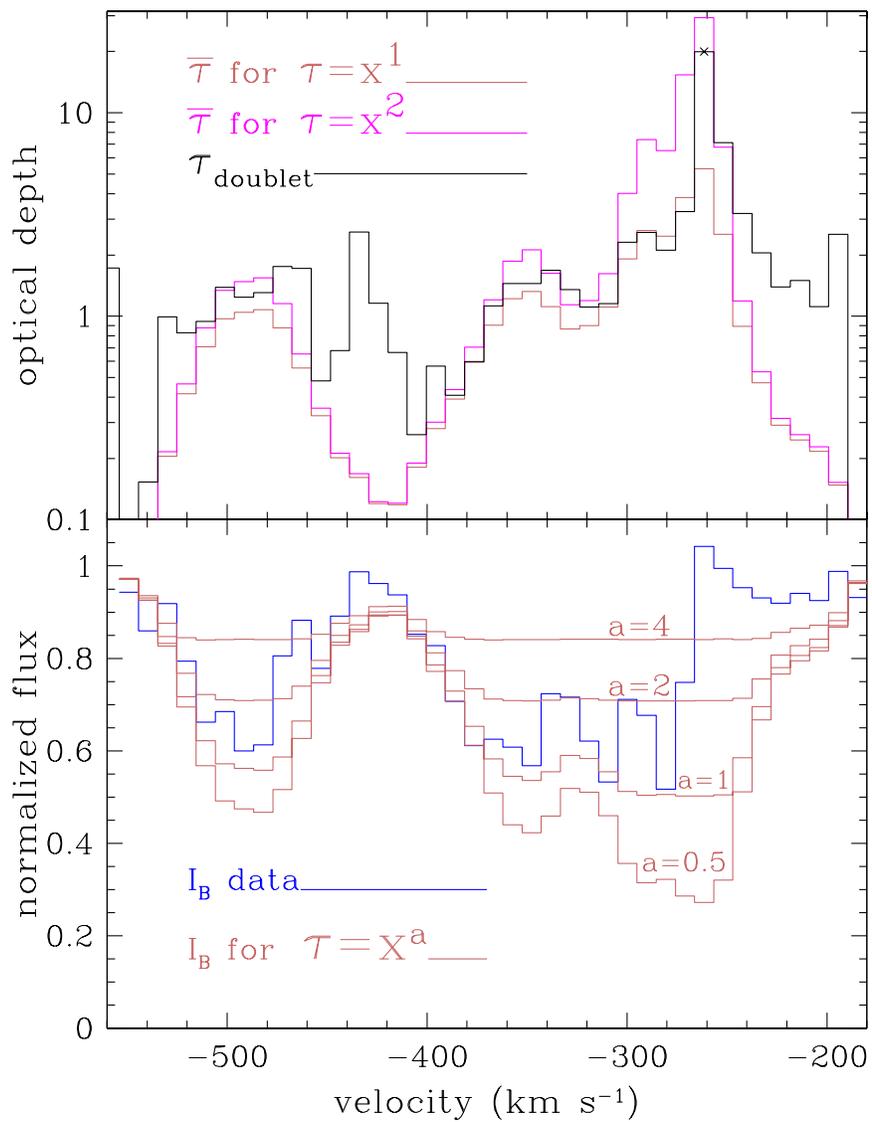,height=16.0cm,width=12.0cm}}
\caption{Optical depth ($\overline{\tau}$) for the four solutions
shown in Fig.\ \ref{fig_o6_renorm} and the $\tau_{\rm{doublet}}$
solution for comparison.  The crosses at the highest
$\tau_{\rm{doublet}}$ values donate lower limits.  It is clear that
the better fit power-law models ($a$=2 and 4) require an enormous
$\overline{\tau}$ for obtaining the fit, while those with more
acceptable $\overline{\tau}$ ($a=0.5$ and 4) do not yield a good fit
for $I_B$ (see Figs. \ref{fig_o6_a1_a4} and \ref{fig_o6_renorm}).  }
\label{fig_n5_tau}
\end{figure}

The situation is quite different for the \nv\ troughs.  Figure
\ref{fig_n5_a1} shows the fitting results for an $a=1$ power-law model
that fit the data quite well.  Examining figure \ref{fig_n5_tau} shows
that power-law models with $1<a<2$ yield good fits except at $v>-260$
\kms. Moreover, the inferred optical depth values are generally in
good agreement with the doublet solution.  The two exceptions are the
expected departure at $v>-260$ \kms\ and at $-400>v>-450$, where the
doublet solution is suspect due to the shallowness of the absorption
data.  

We conclude that by themselves, the \nv\ troughs give  support
to simple inhomogeneous absorber models (even for a model with a
constant gradient).  A reasonable fit is produced with only one
velocity-dependent free parameter (the optical depth), compared to the
partial covering model that requires two free parameters at each
velocity point (optical depth and covering fraction). However,
as we elaborate in the discussion, much of this result can be attributed
to the lower real optical depth on the \nv\ troughs combined with the 
mathematical attributes of the power-law fits.

\clearpage

\clearpage

\subsection{Lyman Series Troughs}

A more stringent test for simple inhomogeneous absorber models is
available by fitting the Lyman series troughs.  In the combined
HST/FUSE spectrum of Mrk~279 we have high-quality data for moderately
deep troughs from \Lya, \Lyb\ and \Lyg.  Fitting three troughs from the
same ion is far more constraining than fitting a doublet.  For a
doublet, we use one of the troughs as a baseline, that is, we construct
the model to give it a perfect fit.  The same model is than used to
fit the other doublet component as described in \S~4.1.  A degree of
freedom (in our case the power-law exponent) may very well suffice to
fit the single remaining trough.  For the three Lyman series troughs, we
fit \Lyb\ in the same way we fitted the red doublet component of
\ovi\ and \nv, but now the same model has to fit both \Lya\ and \Lyg.
To fit the \Lya\ trough we take the derived $\tau_{\rm{max}}$(\Lyb)
 and multiply
it by the expected ratio of optical depths between \Lya\ and \Lyb:
$\vy{\tau}{\Lya}/\vy{\tau}{\Lyb}=(\vy{\lambda}{\Lya}\vy{f}{\Lya})/
(\vy{\lambda}{\Lyb}\vy{f}{\Lyb})$, where $\lambda$ and $f$ 
are the wavelength and oscillator strength for each transition.
We then use this value as  $\tau_{\rm{max}}$(\Lya) for the same power
law model in order to arrive at the expected \Lya\ residual intensity.
The procedure to fit \Lyg\ is similar.  

\begin{figure}
\centerline{\psfig{file=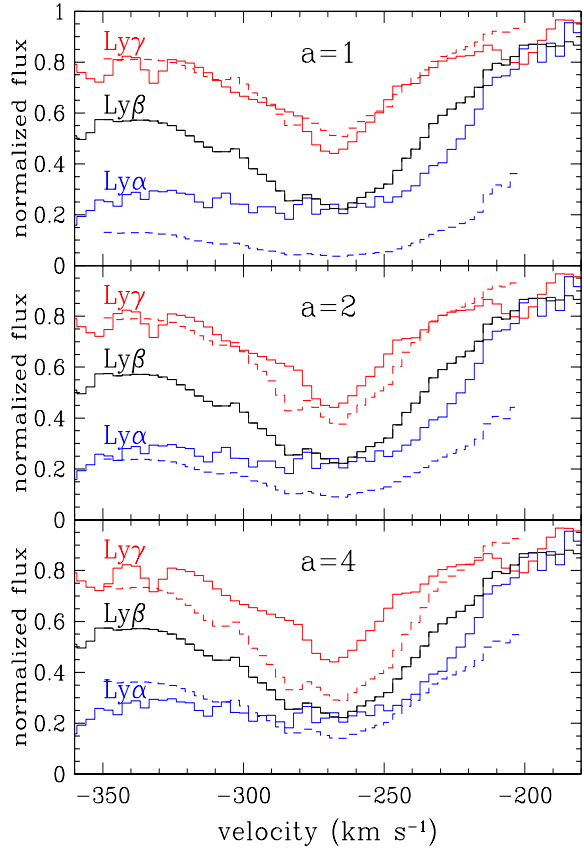,angle=0,height=12.0cm,width=8.0cm}}
\caption{Power-law fits to the Lyman series outflow troughs seen in
Mrk~279 (2003, FUSE and HST spectrum).  The normalized data are plotted
as solid histograms. Errors in the normalized data are dominated by
systematic effects (mainly in the level of the unabsorbed emission
model), which are smaller than 5\% for each curve (Gabel et~al.\
2004). Each panel shows a power-law model ($\tau=x^a$) fit for \Lya\
and \Lyg\ (dashed histograms, color matched to the data).  By
construction the models give a perfect fit to \Lyb. From the range of
plotted models, it is clear that no power-law model can simultaneously
fit \Lya\ and \Lyg.}
\label{lyman_series}
\end{figure}

As described in Scott et~al.\ (2004) and Gabel et~al.\ (2004) the
Lyman series troughs are partially contaminated by non-outflow related
absorption.  Only the absorption in the velocity range $-350<v<-200$
is considered to be purely outflow related in the Lyman series
troughs, and therefore we only fit this region.  The high ionization
CNO doublets are not affected by this unrelated absorption, which allows
us to fit their entire profile.  In figure \ref{lyman_series} we show
the fitting results for three power-law models. While the $a=1$ model
gives a reasonable fit for the \Lyg\ data, it strongly over predicts
the depth of the \Lya\ trough.  A Steeper power law $a=4$ gives an
adequate fit to the \Lya\ trough, but greatly over predicts the depth
of the \Lyg\ trough.  A middle model ($a=2$) gives unsatisfactory fits
for both \Lya\ and \Lyg.  It is clear that simple power-law
inhomogeneous absorber models are not an adequate model for the
Lyman series outflow troughs in Mrk~279.

\section{DISCUSSION}

In \S~4 we attempted to fit the outflow troughs of Mrk~279
with simple power-law inhomogeneous absorber models.
We demonstrated that the \ovi\ troughs cannot be fitted well
with such models.  The main reason for that was the small differences
between $I_B$ and $I_R$ across most of the velocity profile.
The extreme occurrence is in the velocity range $-360<v<-240$  
where $I_B$ and $I_R$ are indistinguishable within the noise limits.
This part of the trough can only be fitted by inhomogeneous models
that will closely resemble the  step-function 
shown in figure~\ref{fig:tau_dist}.  Therefore, the velocity 
range $-360<v<-240$ of the \ovi\ troughs strongly supports the 
physical picture of the partial covering model and advocates for an
optical depth distribution with a sharp edge.

In contrast the \nv\ troughs were well-fitted by a simple
inhomogeneous power-law absorber model.  By itself this is quite a
remarkable fit since we only let the optical depth vary as a function
of velocity, where the partial covering model requires both the
optical depth and the covering fraction to be velocity dependent free
parameters.  Such a situation was found by dKKA for the \siiv\ troughs
in BALQSO~1603+3002.  However, for the Mrk~279 data similar models
could not reasonably fit the \ovi\ troughs nor the Lyman series
troughs.  We should therefore look at the mathematical behavior of the
power-law fits and see whether it can explain the goodness of the \nv\
fit irrespectively of the actual physical model.  We can attribute the
success of the \nv\ fit to the fact that for these troughs $I_B$ is
roughly in between $I_R$ and $I_R^2$. As shown in
figure~\ref{fig:tau_doublet_tau_av}, such occurrence with similar
residual intensities can be reproduced by a shallow power-law model
(e.g., $a=1$) that produces almost identical $\overline{\tau}$ and
$\tau_{\rm{doublet}}$.  It is plausible that the good \nv\ fit is
simply a result of the moderate optical depth for this ion.

As discussed in \S~4.3, the strongest constraints on simple
inhomogeneous absorber models come from the attempt to fit the Lyman
series troughs.  In this case it is evident that even allowing for
complete freedom of both the power-law exponent and the optical depth
as a function of velocity will not yield an acceptable fit.  In
contrast, partial covering models do yield good fits for the Lyman
series troughs (Gabel et~al.\ 2004).  The success of partial covering
models, combined with the failure of the inhomogeneous absorber
models, suggest a sharp edge in the optical depth distribution across
the source.  We note that calculations of synthetic line profiles,
based on wind models with the extended continuum source, support our
notion of the optical depth distribution with a sharp edge (e.g.,
Proga 2003, and references therein).  Partial
covering models are equivalent to a step function distribution of
optical depth across the absorber.  Mathematically, step functions,
gaussians and a power-law distributions all have two free
parameters. Therefore, the comparison between the models stands on
firm grounds.  We conclude that monotonic inhomogeneous absorber
models that do not include a sharp edge in the optical depth
distribution across the source are not an adequate physical model to
explain the trough formation mechanism for the outflow observed in
Mrk~279.

\clearpage


\section*{ACKNOWLEDGMENTS}

This work is based on observations obtained with {\em HST} and {\em
FUSE}, both built and operated by NASA.  Support for this work was
provided by NASA through grants number {\em HST}-AR-9536, {\em
HST}-GO-9688, {\em HST}-GO-9688, from the Space Telescope Science
Institute, which is operated by the Association of Universities for
Research in Astronomy, Inc., under NASA contract NAS5-26555, and
through {\em Chandra} grant 04700532 and by NASA LTSA grant 2001-029.
The National Laboratory for Space Research at Utrecht is supported
financially by NWO, the Netherlands Organization for Scientific
Research. We also thank Yuri Levin for a thorough reading of the
manuscript.



\end{document}